\documentclass{jpsj2}
\title{Resonant Magnetic X-ray Diffraction Study on the
  Triangular Lattice Antiferromagnet GdPd$_2$Al$_3$}

\author{Toshiya \textsc{Inami}, Noriki \textsc{Terada}$^{1}$, 
  Hideaki \textsc{Kitazawa}$^{1}$, and Osamu \textsc{Sakai}$^{1}$}

\inst{
  Synchrotron Radiation Research Unit, Japan Atomic Energy Agency, Hyogo 679-5148, Japan\\
  $^{1}$National Institute for Materials Science, Tsukuba 305-0047,
  Japan}

\abst{Resonant magnetic x-ray diffraction experiments were carried out
  on the stacked triangular lattice antiferromagnet GdPd$_2$Al$_3$.
  The experiments revealed an expected initial collinear c-axis order
  at $T_{\rm N1}$ followed by an additional in-plane order at $T_{\rm
    N2}$, while at the same time we found that the ground state is a
  helically ordered state of a very long incommensurate period of
  approximately 700~{\AA}. The distribution of K-domains was highly
  anisotropic, and the domain with the modulation vector normal to the
  surface of the crystal was ascendant. Low-field magnetization is
  discussed on the basis of the observed incommensurate magnetic
  structure.}

\kword{resonant x-ray diffraction, magnetic x-ray diffraction,
  triangular lattice antiferromagnet}

\begin{document}
\renewcommand{\vec}[1]{\mbox{\boldmath $#1$}}
\maketitle

\section{Introduction} 

Geometrical frustration has occupied a central position in condensed
matter physics for decades\cite{frustration}. Stacked triangular
lattice antiferromagnets were extensively investigated at the early
stages of the history, particularly on the quasi-one-dimensional
hexagonal $ABX_3$ antiferromagnets. Although these magnets do not
posses liquid-like ground states\cite{Tb,GGG,YMn}, which are a
hallmark of highly frustrated magnets, and undergo phase transitions
to long-range ordered states, the novel ordered states and the
distinctive nature of phase transitions found in these magnets
illuminate prominent aspects of various frustration-related
phenomena~\cite{CsCoCl3,CsMnBr3-1,CsMnBr3-2}.

\begin{figure}[tb]
\begin{center}
\includegraphics[width=220pt]{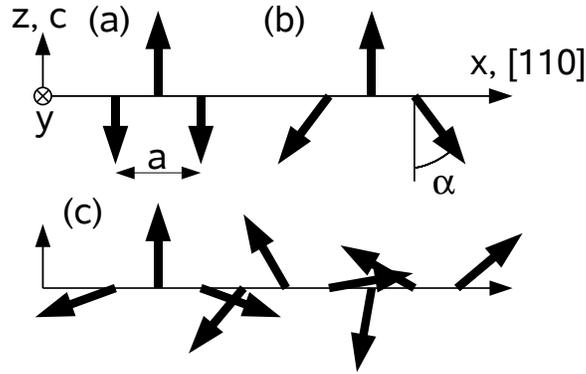}
\end{center}
\caption{
  Magnetic structures of Heisenberg triangular lattice antiferromagnet
  with weak Ising anisotropy. Projection onto the $(1\bar{1}0)$ plane.
  (a) Between $T_{\rm N1}$ and $T_{\rm N2}$ and (b) below $T_{\rm
    N2}$. The canting angle $\alpha$ is 60$^\circ$ for the pure
  Heisenberg case and decreases as the weak Ising anisotropy
  increases. (c) Example of incommensurate structures.}
\label{f1}
\end{figure}

Phase transitions of Heisenberg spins with weak Ising anisotropy on a
triangular lattice are very well investigated\cite{miyashita}. The
characteristic features of these magnets are summarized in the
following three points. (i) At zero field, successive phase
transitions of the $z$- and $xy$-components of the spins occur. As the
temperature decreases, the $z$-component first enters the long-range
ordered state at $T_{\rm N1}$ as shown in Fig.~\ref{f1}(a), and then
the $xy$-components exhibit long-range ordering at $T_{\rm N2}$.
Finally, a slightly distorted 120$^\circ$ structure (Fig.~\ref{f1}(b))
is completed below $T_{\rm N2}$. (ii) Since the canting angle $\alpha$ is
smaller than 60$^\circ$, the total sum of the magnetic moments in a
triangle does not cancel out. Accordingly, a net moment along the
$z$-axis is observed in a single triangular plane. (iii) When the
magnetic field is applied along the $z$-axis, the magnetization is
maintained at a constant value at a certain range of magnetic fields.
The value is exactly one-third of the saturation magnetization, and the
phenomenon is known as the ``one-third (magnetization) plateau''.

The first point, successive phase transitions, was experimentally
confirmed by Clark and Moulton more than 30 years ago in
CsNiCl$_3$\cite{CsNiCl3}, which is one of the above-mentioned $ABX_3$
compounds. However, the predominant antiferromagnetic coupling along
the chain of these compounds inhibits the appearance of the second and
third points, namely a small net moment along the $z$-axis and the
one-third magnetization plateau. The experimental observation of the
latter two points was therefore delayed until the recent discovery
of the prototypical compound GdPd$_2$Al$_3$\cite{kitazawa1}.

GdPd$_2$Al$_3$ crystallizes into a hexagonal structure (space group
$P_6/mmm$; $a$=5.39~{\AA} and $c$=4.19~{\AA})\cite{colineau}. The
magnetic Gd ions ($S$=7/2) are well approximated by Heisenberg spins
and form a stacked triangular lattice. The magnetic properties of this
compound were investigated in detail by Kitazawa and
coworkers\cite{kitazawa1,kitazawa2}. The magnetic susceptibility well
follows the Curie-Weiss law, and the Curie temperature is reported to
be $-$30~K. Hence, the dominant interactions are antiferromagnetic.
The specific heat measurements demonstrated successive phase
transitions at $T_{\rm N1}$ = 16.8~K and $T_{\rm N2}$ = 13.3~K. The
magnetization at low fields indicated a weak ferromagnetic moment
along the $c$-axis. Therefore, it is expected that triangular layers
with small net moments couple ferromagnetically with each other. At
high fields, the one-third magnetization plateau was observed between
6.2~T and 11.8~T, when only the magnetic field parallel to the
$c$-axis was applied.  All of these pieces of evidence are clearly in
line with the criteria of a Heisenberg stacked triangular lattice
antiferromagnet with weak Ising anisotropy. Nevertheless, the magnetic
structures of GdPd$_2$Al$_3$ have not yet been investigated, since
neutron diffraction experiments of Gd compounds are fairly difficult
to perform owing to the large absorption cross section of natural Gd for
neutrons. No microscopic measurements have been reported thus far, with
the exception of a M\"ossbauer spectroscopy experiment, which suggests
a non-collinear magnetic structure of more than two
sublattices\cite{colineau}. The order parameters at $T_{\rm N1}$ and
$T_{\rm N2}$ have not yet been revealed.

In this paper, we carried out resonant magnetic x-ray diffraction
experiments on GdPd$_2$Al$_3$. The temperature dependence of rotated
and unrotated signals with respect to the incident x-ray polarization
unambiguously illustrated successive phase transitions of the $c$-axis
and $ab$-plane components at $T_{\rm N1}$ and $T_{\rm N2}$,
respectively. An unexpected outcome was that the magnetic structure
below $T_{\rm N2}$ was incommensurately modulated. The period was
found to be about 130 times as long as the lattice constant $a$. In
addition, it was found that three domains with respect to the
direction of the modulation wave vector (K-domains) were distributed
extremely unequally. We also discuss the low-field magnetization from
the view point of this long-period incommensurate magnetic structure.

\section{Experimental}

Resonant x-ray diffraction experiments were carried out at beamline
BL22XU in SPring-8. The photon energy was tuned near the $L_2$
absorption edge of Gd (7.930~keV). A single crystal of GdPd$_2$Al$_3$
was grown by the Czochralsky pulling method in a tetra-arc furnace.
The sample was cut into a parallelepiped of
4~$\times$4~$\times$2~mm$^3$ in volume, and a (110) surface was
polished. The sample was attached to the cold head of a conventional
closed-cycle refrigerator, which was mounted on a conventional
four-circle diffractometer with a horizontal scattering plane. We set
the $c$-axis of the sample perpendicular to the scattering plane. The
orientation of the sample was determined using the 110 and 111
reflections, and we confirmed that the angle between the $c$-axis and
the normal vector of the scattering plane was less than 7$^\circ$. The
polarization of the scattered x-rays was analyzed with respect to
whether it was normal ($\sigma$) or parallel ($\pi$) to the scattering
plane using the 006 reflection of a pyrolytic graphite crystal.
Since the incident polarization was parallel to the scattering plane
($\pi$), the scattered x-rays were separated into rotated ($\pi$ to
$\sigma^\prime$) and unrotated ($\pi$ to $\pi^\prime$) channels, where
the prime sign represents scattered x-rays. In order to prevent the
beam from heating the sample surface, we reduced the incident photon
flux by a factor of 7. The experiments were carried out twice. In the
first experiment, the mosaic width of the 110 reflection of the sample
was about 0.07$^\circ$ full width at half maximum (FWHM), indicating
the high quality of the crystal. Prior to the second experiment, the
sample surface was carefully polished again, and the mosaic width was
improved to 0.015$^\circ$ FWHM. The data provided in the rest of this
paper were obtained from the repolished sample, unless explicitly
specified otherwise.

\section{Results}

\begin{figure}[tb]
\begin{center}
\includegraphics[width=220pt]{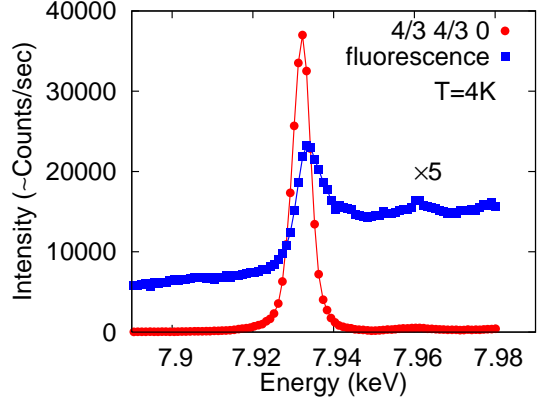}
\end{center}
\caption{
  (color online) Energy dependence of the peak intensity of a magnetic
  Bragg reflection at ($\frac{4}{3},\frac{4}{3},0$). The fluorescence
  spectrum is also shown. Absorption correction is not performed.  The
  diffraction intensity resonantly increases near the Gd $L_2$
  absorption edge 7.931~keV.}
\label{f2}
\end{figure}

First, we show the peak intensity of the magnetic reflection
$\frac{4}{3}\frac{4}{3}0$ as a function of photon energy at 4~K, as
well as the fluorescence spectrum, in Fig.~\ref{f2}. Absorption
correction was not performed. The fluorescence spectrum indicates that
the Gd $L_2$ absorption edge is 7.931~keV. The diffraction intensity
was very weak far from the $L_2$ absorption edge, while an enormous
resonant enhancement of the intensity was observed near the main edge,
implying that the resonance is ascribed to electric dipole (E1)
transitions ($2p\rightarrow5d$). Subsequent measurements were
performed at the peak energy (7.932~keV). We confirmed that the
modulation wave vector $\vec q_{\rm M}$ is only
($\frac{1}{3},\frac{1}{3},0$) in the first experiment. Reciprocal
lattice scans from ($\frac{2}{3},\frac{2}{3},0$) to
($\frac{2}{3},\frac{2}{3},1$) and from (0.4,0.4,0) to (1.7,1.7,0) yielded
four magnetic reflections at ($\frac{2}{3},\frac{2}{3},0$),
($\frac{2}{3},\frac{2}{3},1$), ($\frac{4}{3},\frac{4}{3},0$), and
($\frac{5}{3},\frac{5}{3},0$). No other modulation wave vectors were
detected. We also surveyed an entire Brillouin zone by using an area
detector and confirmed the above result\cite{inami}.

\begin{figure}[tb]
\begin{center}
\includegraphics[width=220pt]{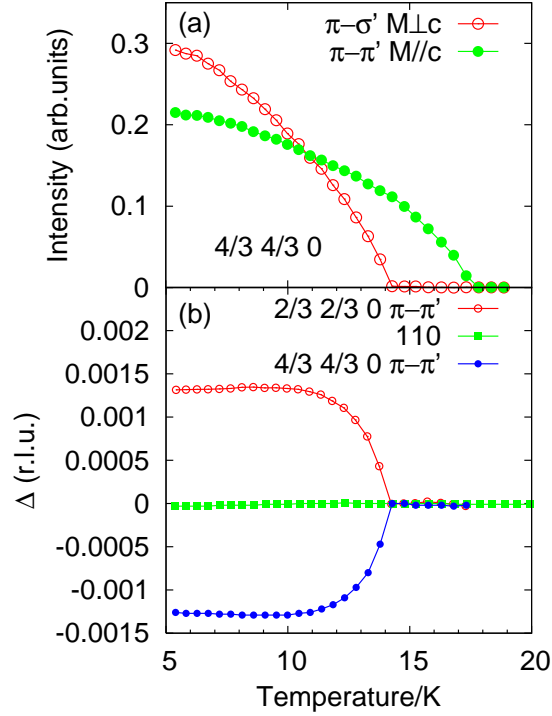}
\end{center}
\caption{
  (color online) (a) Integrated intensity of the rotated
  ($\pi$-$\sigma^\prime$) and unrotated ($\pi$-$\pi^\prime$) channels
  of the $\frac{4}{3}\frac{4}{3}0$ magnetic reflection as a function
  of temperature. The rotated and unrotated channels sense the
  magnetic moments normal and parallel to the $c$-axis, respectively.
  See the text for details. (b) Deviations of the peak position from
  the commensurate position ($\Delta$) for the 110,
  $\frac{4}{3}\frac{4}{3}0$, and $\frac{2}{3}\frac{2}{3}0$ reflections.
  The unit is a reduced lattice unit. The unrotated channels are
  depicted for the magnetic reflections. The deviations of the
  magnetic reflections clearly develop below $T_{\rm N2}$.}
\label{f3}
\end{figure}

The main result was obtained from the temperature dependence of the
$\frac{4}{3}\frac{4}{3}0$ reflection. In Fig.~\ref{f3}(a), the
diffraction intensity integrated along the radial ($\theta$-2$\theta$)
direction is shown as a function of temperature for both the
$\pi$-$\pi^\prime$ and $\pi$-$\sigma^\prime$ channels. The data
unambiguously illustrate that the $\pi$-$\pi^\prime$ channel appears
below 17.3~K ($T_{\rm N1}$), whereas the $\pi$-$\sigma^\prime$ channel
is observed only below 14.3~K ($T_{\rm N2}$). The slight discrepancies
(about 1~K) between the observed transition temperatures and the
values cited in the literature are probably due to radiation heating
of the thermometer. In the intermediate phase between $T_{\rm N1}$ and
$T_{\rm N2}$, only the scattering process from $\pi$ to $\pi^\prime$
is allowed.

Here we briefly describe the resonant x-ray scattering amplitude
$f_{\rm res}$ of Gd ions. According to ref.~\citen{blume}, $f_{\rm
  res}$ is proportional to
\begin{equation}
  \label{eq:1}
f_{\rm res} \propto C_0 {\vec\epsilon}_{\rm f}^*\cdot{\vec\epsilon}_{\rm i}
+ i C_1 ({\vec\epsilon}_{\rm f}^*\times{\vec\epsilon}_{\rm i})\cdot{\vec m}+ 
C_2\,{\vec\epsilon}^\dag_{\rm f}\,O\,{\vec\epsilon}_{\rm i},
\end{equation}
where $C_0$, $C_1$, and $C_2$ are energy-dependent constants, $\vec m$
is the magnetic moment, and ${\vec \epsilon}_{\rm i}$ and ${\vec
  \epsilon}_{\rm f}$ are the polarization vectors of the incident and
scattered x-rays, respectively. The symmetric second-rank tensor $O$
describes the anisotropy of the Gd 5$d$ orbital caused by the
anisotropic crystal environment. We neglect a small magnetic
contribution in $O$, which is proportional to $m^2$\cite{blume,doon}.
The first term is an ordinary anomalous scattering factor. The second
term corresponds to magnetic scattering, and the last term causes the
anisotropy of the tensor of susceptibility (ATS) scattering. Since all
Gd ions are equivalent in GdPd$_2$Al$_3$, the first term does not
contribute to super-lattice reflections. It is also expected that the
last term, ATS scattering, does not produce any intensity at
super-lattice positions. This is due to the fact that Gd ions have no
quadrapole moments. Hence, it is reasonable to consider that the
magnetic ordering does not introduce anisotropic lattice distortions
around the Gd ions. Therefore, the second term, magnetic scattering,
is the only term to be considered in GdPd$_2$Al$_3$.

As already mentioned, $\pi$ polarization is parallel to the scattering
plane. In addition, we set the $c$-axis of the crystal perpendicular
to the scattering plane. Since the resonant x-ray magnetic scattering
amplitude includes an outer product of ${\vec \epsilon}_{\rm i}$ and
${\vec \epsilon}_{\rm f}$, the $\pi$-$\pi^\prime$ channel observes
magnetic moments perpendicular to the scattering plane, which are the
$c$-axis component of the magnetic moments. In contrast, the
scattering amplitude of the $\pi$-$\sigma^\prime$ channel is
proportional to magnetic moments parallel to the incident x-rays,
which are the magnetic moments in the $ab$-plane. By applying this
characteristic polarization dependence of resonant magnetic x-ray
diffraction, the temperature dependence of the magnetic reflection
shown in Fig.~\ref{f3}(a) directly leads to the conclusion that the
phase transition at $T_{\rm N1}$ is an ordering of the $z$-component
of the magnetic moments and that the $xy$-components are paramagnetic
until the temperature drops below $T_{\rm N2}$, in good accordance
with the expected behavior of Heisenberg triangular lattice
antiferromagnets with weak Ising anisotropy.

A minute exploration of the data, however, detects unexpected behavior
as well. Firstly, we found that the transition at $T_{\rm N2}$ is a
commensurate-to-incommensurate transition. The peak positions of the
$\frac{2}{3}\frac{2}{3}0$ and $\frac{4}{3}\frac{4}{3}0$ magnetic
reflections moved slightly along the radial ($\theta$-2$\theta$)
direction below $T_{\rm N2}$. We show the deviations of the peak
position from the commensurate position ($\Delta$) for the
$\pi$-$\pi^\prime$ channels of two magnetic reflections as a function
of temperature in Fig.~\ref{f3}(b), where we denote the wave vectors
of the magnetic reflections as $q=(h+\Delta,h+\Delta,0)$ and
$h=\frac{2}{3}$ or $\frac{4}{3}$. We also show $\Delta$ for the 110
lattice reflection as a reference. The commensurate positions were
estimated at 15~K. The deviations for the $\frac{2}{3}\frac{2}{3}0$
and $\frac{4}{3}\frac{4}{3}0$ reflections developed positively and
negatively below $T_{\rm N2}$ like an order parameter of a
second-order phase transition, respectively, whereas the deviation for
the 110 reflection remained zero throughout the entire range of
temperature in the measurements. The magnetic modulation wave vector
$\vec q_{\rm M}$ is thus expressed as
$(\frac{1}{3}-\delta,\frac{1}{3}-\delta,0)$. $\delta$ was evaluated to
be 0.0013 at 5~K. About 130 lattice points are therefore included in
the period $\frac{1/3}{\delta}$, which is approximately 700~\AA. An
example of incommensurate magnetic structures is shown in
Fig.~\ref{f1}(c).

\begin{figure}[tb]
\begin{center}
\includegraphics[width=220pt]{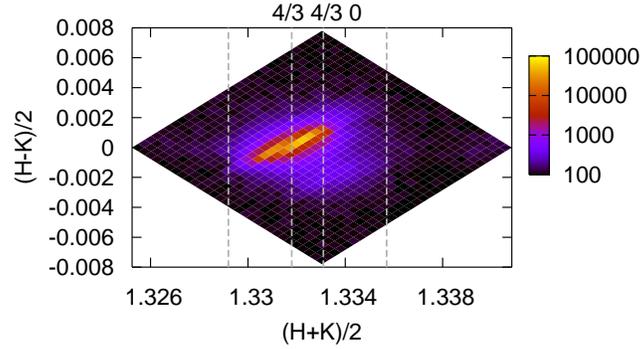}
\end{center}
\caption{ (color online) Intensity map on the ($H$,$K$,0) plane for
  the $\frac{4}{3}\frac{4}{3}0$ reflection at 5.8~K. The intensity is
  plotted on a logarithmic scale, and the units are counts per second.
  Vertical broken lines indicate the positions of $(H+K)$/2 =
  $4/3-3\delta$, $4/3-\delta$, 4/3, and $4/3+2\delta$, from left to
  right. A very weak peak might exist at the lower right of the strong
  peak. Higher harmonic satellites (at 2$q_{\rm M}$ and 3$q_{\rm M}$)
  were not observed.}
\label{f4}
\end{figure}

Secondly, the distribution of domains was not uniform. In the
incommensurate phase, there were three domains with respect to the
direction of the modulation vectors (the so-called K-domains). For
instance, the $\frac{1}{3}\frac{1}{3}0$ reflection splits into three
Bragg points at $(\frac{1}{3}-\delta,\frac{1}{3}-\delta,0)$,
$(\frac{1}{3}-\delta,\frac{1}{3}+2\delta,0)$, and
$(\frac{1}{3}+2\delta,\frac{1}{3}-\delta,0)$. We measured the
intensity map on the ($H,K,0$) reciprocal-lattice plane around the
$\frac{4}{3}\frac{4}{3}0$ reflection at 5.8~K. The result is depicted
in Fig.~\ref{f4}. There should be three domains. However, only one
domain is visible. Although a very weak domain might exist at
$(\frac{4}{3}-\delta,\frac{4}{3}+2\delta,0)$, most of the intensity
concentrates at the primary peak. The modulation vector of the
strongest domain was normal to the crystal surface. We believe that
residual strains introduced by the polishing of the (110) surface
select the observed domain. In the first experiment, the domain
distribution was also quite anisotropic. However, we observed another
domain rather clearly, and hence we consider that repolishing reduced
the volume of minor domains. Figure~\ref{f4} also illustrates that
there are no higher harmonic satellites. The absence of a peak at
2$q_{\rm M}$ suggests that the $m^2$ term in eq.~(\ref{eq:1}) is
actually negligible in GdPd$_2$Al$_3$~\cite{doon}. No observable
intensity at 3$q_{\rm M}$ corresponds to a small squaring up of the
modulation, indicating that the magnetic structure is described only
by the fundamental wave vector $q_{\rm M}$.

The residual strains might affect the direction of the magnetic
moments. Domains with respect to the direction of magnetic moments are
referred to as S-domains. In GdPd$_2$Al$_3$, the magnetic moments
share a single $a^*c$-plane in an S-domain. Since three equivalent
$a^*c$-planes exist in a hexagonal lattice, there are three S-domains.
The distribution of S-domains can be estimated from the ratio of the
intensity of the $\pi$-$\pi^\prime$ channel to that of the
$\pi$-$\sigma^\prime$ channel if the canting angle $\alpha$ (see
Fig.~\ref{f1}(b)) is known, where we assume the commensurate structure
for the sake of simplicity. $\alpha$ is evaluated to be 59$^\circ$
from the weak ferromagnetic moment along the $c$-axis at 0~T.
Although this value requires a slight correction, as seen in the
discussion section, it is certain that $\alpha$ is very close to
60$^\circ$ and hence we assume $\alpha$=60$^\circ$. The ratios of
$\pi$-$\sigma^\prime$ to $\pi$-$\pi^\prime$ reached constant values
1.36 and 5.1 at low temperatures for the $\frac{4}{3}\frac{4}{3}0$ and
$\frac{2}{3}\frac{2}{3}0$ reflections, respectively. This result leads
to the conclusion that about 70\% of domains have a spin plane normal to
the modulation vector. In contrast, the equal distribution of
S-domains provided a reasonable account in the first experiment.

\section{Discussion}

\begin{figure}[tb]
\begin{center}
\includegraphics[width=220pt]{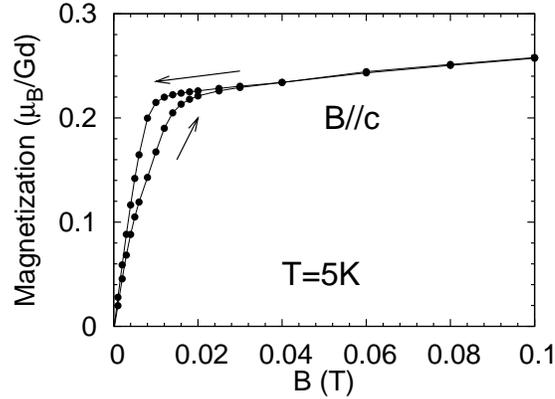}
\end{center}
\caption{Low-field magnetization at 5~K when magnetic field parallel
  to the $c$-axis is applied. (Data taken from
  ref.~\citen{kitazawa1}.)}
\label{f5}
\end{figure}

\subsection{Magnetic structure}

Throughout these experiments, we found that there is only one magnetic
wave vector $\vec{q}_{\rm
  M}=(\frac{1}{3}-\delta,\frac{1}{3}-\delta,0)$ in GdPd$_2$Al$_3$. In
addition, we observed the magnetic moments parallel to the $c$-axis
${\vec M}_\parallel$ and the magnetic moments perpendicular to the
$c$-axis ${\vec M}_\perp$. Therefore the magnetic moment {\vec M}
below $T_{\rm N2}$ at the position ${\vec r}$ is given by
\[
{\vec M(\vec{r})} = {\vec M}_\parallel \sin (\vec{q}_{\rm M}\vec{r}+\phi_0)
+ {\vec M}_\perp \sin (\vec{q}_{\rm M}\vec{r}+ \phi_0 + \phi).
\]
At $\phi=\pi/2$, the magnetic structure is a helically modulated
structure such as that shown in Fig.~\ref{f1}(c). On the other hand,
when $\phi=0$, the magnetic structure is a sinusoidally modulated one,
which inclines from the $c$-axis by $\theta_c=\tan^{-1}(|{\vec
  M}_\perp|/|{\vec M}_\parallel|)$. Although it is not easy to
determine the value of $\phi$ experimentally\cite{phi}, it can be
reasonably inferred that $\phi$ is $\pi/2$. Firstly, a sinusoidally
modulated structure is unfavorable with regard to entropy at low
temperatures. Fully polarized magnetic moments are expected for such
localized magnets. Secondly, an inclined structure is not stable. The
origin of the anisotropy must be a dipole-dipole interaction since the
Gd ion has no orbital moment. Owing to the ferromagnetic coupling
between the layers along the $c$-axis, the dipole-dipole interaction
behaves as effective Ising anisotropy. If a sinusoidally modulated
structure is realized, $\theta_c$ tends to be zero. Accordingly, we
conclude that the magnetic structure below $T_{\rm N2}$ is a helically
and incommensurately modulated structure.

\subsection{Incommensurate modulation}

Normally, an incommensurate structure is ascribed to competition
between the nearest-neighbor interaction and further-neighbor
interactions\cite{screw}. Long-range RKKY interactions are relevant
for the metallic GdPd$_2$Al$_3$. Hence, further-neighbor interactions
can be comparable in magnitude to the nearest-neighbor interaction. It
is reasonable to assume that the observed incommensurate structure is
stabilized by second- and third-neighbor antiferromagnetic
interactions. These isotropic interactions, however, do not explain
why the intermediate phase is commensurate. The concomitant
development of the incommensurability $\delta$ and the in-plane
component $M_\perp$ below $T_{\rm N2}$ shown in Fig.~\ref{f3} strongly
suggests that $M_\perp$ is responsible for the incommensurate
structure through anisotropic interactions.

In-plane incommensurate magnetic structures in stacked (hexagonal)
triangular lattice antiferromagnets were observed in
RbFeCl$_3$\cite{wada}. The phase transitions in RbFeCl$_3$ are quite
peculiar. There are two (linearly-polarized) incommensurate structures
at high temperatures, and the low-temperature structure is a
commensurate coplanar one. These incommensurate magnetic structures
and the novel ordering process are well interpreted by introducing
dipole-dipole interactions between the in-plane magnetic
moments\cite{shiba}. The exchange energy has a minimum at the K point
($=(\frac{2\pi}{3a},\frac{2\pi}{3a},0)$), while the dipole-dipole
interactions between the in-plane magnetic moments prefer other
ordering wave vectors. Hence, stable points appear near the K point,
and incommensurate structures are realized. In contrast, dipole-dipole
interactions between the magnetic moments parallel to the $c$-axis are
stable at the K point. Therefore, dipole-dipole interactions provide a
reasonable account of incommensurate-commensurate transitions at
$T_{\rm N2}$.  Detailed calculations will be published elsewhere.

\subsection{Low-field magnetization}

In the rest of the paper, we discuss the low-field magnetization of
GdPd$_2$Al$_3$. As shown in Fig.~\ref{f5}, it appears that the
magnetization along the $c$-axis has a small ferromagnetic moment.
The extrapolation of the magnetization curve above 0.02~T intersects
about 0.22~$\mu_{\rm B}$/Gd at 0~T. If the magnetic structure is
commensurate, this weak net moment is in good agreement with the
expectation that a distorted 120$^\circ$ structure due to the weak
Ising anisotropy gives rise to a small ferromagnetic moment in a
single triangular plane since the triangular planes couple
ferromagnetically with each other, as experimentally confirmed by the
zero $z$-component of the modulation wave vector. However, the
observed magnetic structure is incommensurate. Since the magnetic
structure is very close to the 120$^\circ$ structure, a local triangle
of spins may have net moment, for instance, along the $+z$ direction.
However, at a position a half-period away, the net moment of the
triangle points to the $-z$ direction. Therefore, the triangular plane
has no ferromagnetic moment. This discrepancy can be resolved by
inserting domain walls. If the spins rotate clockwise in the first
half-period of the incommensurate structure and rotate
counterclockwise in the next half-period, the total magnetic moments
in all half-period domains point in the same direction. It is likely
that this domain structure is stable in the presence of magnetic
fields. In fact, this domain structure can be regarded as a fan
structure, which was introduced by Nagamiya {\it et al.}\cite{nagamiya}
in the helical-to-fan transition of a helimagnet under magnetic
fields. The transition field to the fan structure is considered to be
low, since the density of the domain walls is extremely low owing to
the long incommensurate period. As the magnetic field increases, the
amplitude of the incommensurate structure decreases, and eventually
the commensurate structure is achieved.

\section{Conclusions}

We investigated the magnetic structures and phase transitions of the
Heisenberg triangular lattice antiferromagnet GdPd$_2$Al$_3$ by means
of resonant magnetic x-ray diffraction. Utilizing the characteristic
polarization dependence of resonant magnetic x-ray diffraction, we
revealed that only the $c$-axis component of the magnetic moments
exhibits long-range ordering between $T_{\rm N1}$ and $T_{\rm N2}$ and
that the helical structure is completed by the ordering of the
$ab$-plane components below $T_{\rm N2}$. The magnetic structure below
$T_{\rm N2}$ was found to be incommensurate and is most likely
stabilized by the dipole-dipole interactions between the in-plane
components of the magnetic moments. We also found that both the
K-domains and the S-domains were not evenly distributed. It is likely
that the residual strains caused by polishing the surface produce
these asymmetric distributions. The low-field magnetizations were
discussed from the viewpoint of the observed incommensurate magnetic
structures, and we infer that the fan structure is stabilized in the
magnetic field.

\section*{Acknowledgment}

We would like to thank Professor Seiji Miyashita for helpful
suggestions. This work was supported by Grant-in-Aid for Scientific
Research C (19540386) and partly supported by Grant-in-Aid for
Scientific Research on priority Areas ``High Field Spin Science in 100
T'' (No.  451) from the Ministry of Education, Culture, Sports,
Science and Technology (MEXT).


\begin{thebibliography}{99} 
\bibitem{frustration} A. P. Ramirez: in \textit{Handbook of Magnetic
    Materials}, ed. K. H. J. Buschow (Elsevier Science B. V.,
  Amsterdam, 2001) 423.
\bibitem{Tb} J. S. Gardner, S. R. Dunsiger, B. D. Gaulin,
  M. J. P. Gingras,  J. E. Greedan, R. F. Kiefl, M. D. Lumsden,
  W. A. MacFarlane, N. P. Raju, J. E. Sonier, I. Swainson, and Z. Tun:
  Phys. Rev. Lett. \textbf{82} (1999) 1012.
\bibitem{GGG} O. A. Petrenko, C. Ritter, M. Yethiraj, and D. McK Paul:
  Phys. Rev. Lett. \textbf{80} (1998) 4570.
\bibitem{YMn} R. Ballou, E. Leli\`evre-Berna, and B. F{\aa}k:
  Phys. Rev. Lett. \textbf{76} (1996) 2125.
\bibitem{CsCoCl3} M. Mekata and K. Adachi: J. Phys. Soc. Jpn. \textbf{44}
  (1978) 806.
\bibitem{CsMnBr3-1} Y. Ajiro, T. Nakashima, Y. Unno, H. Kadowaki,
  M. Mekata, and N. Achiwa: J. Phys. Soc. Jpn. \textbf{57} (1988) 2648.
\bibitem{CsMnBr3-2} B. D. Gaulin, T. E. Mason, and M. F. Collins:
  Phys. Rev. Lett. \textbf{62} (1989) 1380.
\bibitem{miyashita} S. Miyashita: J. Phys. Soc. Jpn. \textbf{55}
  (1986) 3605.
\bibitem{CsNiCl3} R. H. Clark and W. G. Moulton: Phys. Rev. B
  \textbf{5} (1972) 788.
\bibitem{kitazawa1} H. Kitazawa, H. Suzuki, H. Abe, J. Tang, and G.
  Kido: Physica B \textbf{259-261} (1999) 890.
\bibitem{colineau} E. Colineau, J. P. Sanchez, J. Rebizant, and J. M.
  Winand: Solid State Commun. \textbf{92} (1994) 915.
\bibitem{kitazawa2} H. Kitazawa, K. Hashi, H. Abe, N. Tsujii, and G.
  Kido: Physica B \textbf{294-295} (2001) 221.
\bibitem{inami} T. Inami, H. Toyokawa, N. Terada, and H. Kitazawa:
  J. Phys.: Conf. Ser. \textbf{150} (2009) 042069.
\bibitem{blume} M. Blume: in \textit{Resonant Anomalous X-Ray
    Scattering}, ed. G. Materlik, C. J. Sparks, and K. Fischer
  (Elsevier Science B. V., Amsterdam, 1994) 495.
\bibitem{doon} D. Gibbs, D. R. Harshman, E. D. Isaacs, 
  D. B. McWhan, D. Mills, and C. Vettier:
  Phys. Rev. Lett. \textbf{61} (1988) 1241.
\bibitem{phi} In principle, the phase $\phi$ can be estimated through
  diffraction experiments if a single domain is obtained. A helically
  modulated structure converts linearly polarized incident x-rays into
  elliptically polarized x-rays. If incoming x-rays are circularly
  polarized, the scattering amplitude depends on the helicity of the
  x-rays\cite{ho}. In contrast, a sinusoidally modulated structure
  does not generate elliptically polarized x-rays from linearly
  polarized x-rays, and the helicity of incident x-rays does not
  affect the scattering amplitude. However, as yet no adequate method
  exists for obtaining a single domain for a metallic sample. Hence, it
  is difficult to determine the value of $\phi$ in GdPd$_2$Al$_3$.
\bibitem{screw} E. Rastelli, A. Tassi, and L. Reatto: Physica B
  \textbf{97} (1979) 1.
\bibitem{wada} N. Wada, K. Ubukoshi, and K. Hirakawa: J. Phys. Soc.
  Jpn. \textbf{51} (1982) 2833.
\bibitem{shiba} H. Shiba and N. Suzuki: J. Phys. Soc. Jpn. \textbf{52}
  (1983) 1382.
\bibitem{nagamiya} T. Nagamiya, K. Nagata, and Y Kitano: Prog. Theor.
  Phys. \textbf{27} (1962) 1253.
\bibitem{ho} J. C. Lang, D. R. Lee, D. Haskel, and G. Srager: J. Appl.
  Phys. \textbf{95} (2004) 6537.
 \end{thebibliography}
\end{document}